\documentclass[12pt,a4paper,dvips]{article}
\usepackage{a4p}
\usepackage{cite,mcite}
\usepackage{graphicx}
\usepackage{physics }
\usepackage{l3_title,ifthen,Lep}
\journalname{Phys. Lett. B}
\date{January 30, 2001}
\preprint{2001-012}
\newlength{\capindent}
\newlength{\capwidth}
\newlength{\figwidth}
\newcommand{\icaption}[2][!*!,!]{\hspace*{\capindent}%
  \begin{minipage}{\capwidth}
    \ifthenelse{\equal{#1}{!*!,!}}%
      {\caption{#2}}%
      {\caption[#1]{#2}}
  \end{minipage}}

\def\gg    {$\gamma \gamma$} 
\def\gge   {$\gamma ^* \gamma ^*$} 

\def\sggh    {$\sigma(\gamma \gamma \rightarrow$ {\sl hadrons})}

\def\sgg    {$\sigma _{\gamma \gamma }$}
\def\sgge    {\sigma _{\gamma \gamma }}
\def\csee {$ \sigma $(\ee \ra {} \ee {\sl hadrons})} 
\def\seeh {$\Delta \sigma ($\ee \ra {} \ee {\sl hadrons})}
\def\seehe { \Delta \sigma ( \ee \ra   \ee {\sl hadrons})}
\def\dseeh {$\seehe /\Delta W_{\gamma \gamma }$} 

\def\eett {  \ee \ra {}  \ee $\tau ^+ \tau ^-$}

\def\see {$\Delta \sigma _{\ee} $}

\def\Lgg  {$\cal{L}_{\gamma \gamma }$}

\def\Wgge {W_{\gamma \gamma}}
\def\Wgg {$W_{\gamma \gamma }$}
\def\Wvis {$W_{\mathrm{vis}}$}

\def\Wvise {W_{\mathrm{vis}}}
\def\Q2 {$Q^2$}
\def\q2 {$q^2$}
\begin{document}
\bibliographystyle{l3style}
\begin{titlepage}
\title{ Total Cross Section   
        in {\boldmath $ \gamma \gamma $} Collisions at LEP}
\author{The L3 Collaboration}
%
% The abstract
%
%
% The abstract
%
\begin{abstract}
The reaction 
$\rm{e}^{+} \rm{e}^{-} \rightarrow \rm{e}^{+} \rm{e}^{-} \gamma ^{*}  \gamma ^{*}  \rightarrow 
 \rm{e}^{+} \rm{e}^{-}$ {\sl hadrons} for quasi-real photons
is studied  using data from $ \sqrt {s} = 183 \GeV $ 
     up to 202 \GeV .
 Results on the total cross sections 
 $\sigma(\rm{e}^{+} \rm{e}^{-} \rightarrow 
 \rm{e}^{+} \rm{e}^{-} $ {\sl hadrons}) and  
$\sigma (\gamma\gamma \rightarrow $ {\sl hadrons}) are given 
for the two-photon  centre-of-mass energies
 5  GeV $\leq W_{\gamma\gamma} \leq$ 185 GeV.
  The total
  cross section of two real photons is  
   described by a  Regge parametrisation. We observe
 a steeper rise with the two-photon centre-of-mass energy as compared to the
hadron-hadron and the photon-proton cross sections. The data are 
also compared to the expectations of different theoretical models.

\end{abstract}

\vspace*{10mm}
%\centerline{\it { The figures are the old ones, they will be updated soon.}}
%
% Adds "To be submitted to ..." or "Submitted to ...", if relevant
%
 \submitted

%\vspace*{10mm}
%\centerline{Submitted to the}
%\centerline{\it XXXth International Conference on High Energy Physics}
%\centerline{Osaka, Japan, July 27 - August 2, 2000. }
 \end{titlepage}
%
%%%%%%%%%%%%%%%%%%%%%%%%%%%%%%%%%%%%%%%%%%%%%%%%%%%%%%%%%%%%%%%%%%%%%%%%%%%%%%%
% Introduction
%%%%%%%%%%%%%%%%%%%%%%%%%%%%%%%%%%%%%%%%%%%%%%%%%%%%%%%%%%%%%%%%%%%%%%%%%%%%%%%
%
\section{Introduction}
At high centre-of-mass energies, \rts , the two-photon process 
\ee \ra {} \ee \gge \ra {} \ee {\sl hadrons} is 
a copious source of hadron production.
 Most of the initial energy
is taken by the scattered 
electrons\footnote{Electron stands for electron and positron throughout this
paper.}. As  
 their scattering angle is low,
 they often go  undetected. 
 The hadron system has, typically, a low mass compared to \rts. 
A large fraction of the hadrons escape detection, due to the Lorentz boost of the
\gg {} system and to the 
large diffractive cross section producing hadrons at small polar angles, 
where the detector acceptance is limited.
 For these reasons, the measured visible mass, \Wvis ,
 is  less than 
 the two  photon effective mass, \Wgg.
  \par 
 In this paper we analyse  only data
where the scattered electrons are not
detected. 
New results on total cross sections \csee
{} are presented,  using
 data  collected with the L3 detector \cite{L3} for
 a total integrated luminosity of 51.4 \pb {}   
 at \rts = 183 \GeV , 171.8 \pb {} 
 at \rts = 189 \GeV {} 
and 220.8 \pb {} 
 at \rts = 192, 196, 200, 202 \GeV .
\par  
 The  two-photon cross section \sggh {}
is  derived in the  
interval $5 \GeV  \leq W_{\gamma \gamma} \leq  185 \GeV $, while
the analysis  of the data taken at \rts = 133 \GeV {} and 161 \GeV \cite{had161} 
covered only the interval $5 \GeV \leq W_{\gamma \gamma} \leq 75 \GeV $.

%%%%%%%%%%%%%%%%%%%%%%%%%%%%%%%%%%%%%%%%%%%%%%%%%%%%%%%%%%%%%%%%%%%%%%%
% L3
%%%%%%%%%%%%%%%%%%%%%%%%%%%%%%%%%%%%%%%%%%%%%%%%%%%%%%%%%%%%%%%%%%%%%%%
\section{ Measurement of cross sections}

%%%%%%%%%%%%%%%%%%%%%%%%%%%%%%%%%%%%%%%%%%%%%%%%%%%%%%%%%%%%%%%%%%%%%%%%%%
% events selection
%%%%%%%%%%%%%%%%%%%%%%%%%%%%%%%%%%%%%%%%%%%%%%%%%%%%%%%%%%%%%%%%%%%%%%%%%%
\subsection{Monte Carlo simulation}
The \ee \ra {} \ee \gge \ra {} \ee {\sl hadrons} processes are generated with the 
   PHOJET 
\cite{Engel} and PYTHIA  \cite{PYTHIA} 
event generators.
For the annihilation  process 
\ee $\rightarrow$ \qqbar ($\gamma $),
  PYTHIA  is used. 
  KORALZ \cite{KORALZ} is used for \ee $\rightarrow \tau^{+} \tau^{-}(\gamma )$
and KORALW \cite{KORALW} for \ee $\rightarrow \rm{W^{+} W^{-}}$.
The  \ee \ra {} \ee $\tau^{+} \tau^{-}$ channel 
is generated by   DIAG36 \cite{DIAG36}.  
 The Monte Carlo events 
are simulated in the L3 detector using the GEANT \cite{GEANT}
and GHEISHA \cite{GEISHA} programs
and 
passed through the same reconstruction program as the data.
Time dependent detector inefficiencies, as monitored during the data taking
period, are also simulated.
\subsection{Event selection}
\label{sel}

The analysis  is   based on 
the central tracking system,  
  the high resolution electromagnetic
calorimeter, the  hadron calorimeter and  
 the luminosity monitor.
\par
Two-photon  events  are collected 
predominantly by the track triggers \cite{trigger}.
 The trigger efficiency 
 is    studied separately for each data sample  
by comparing  the number of events accepted by 
 the track trigger and the  
calorimetric energy trigger. The efficiencies of higher level triggers are measured using
 prescaled  events. The trigger efficiency  
 increases from
 80 \% at $\Wvise = 5 \GeV $  to  94 \%  above 80 \GeV.

\par
\begin{figure}[htbp]
 \begin{tabular}{ll}
  \includegraphics[width=0.45\textwidth ]{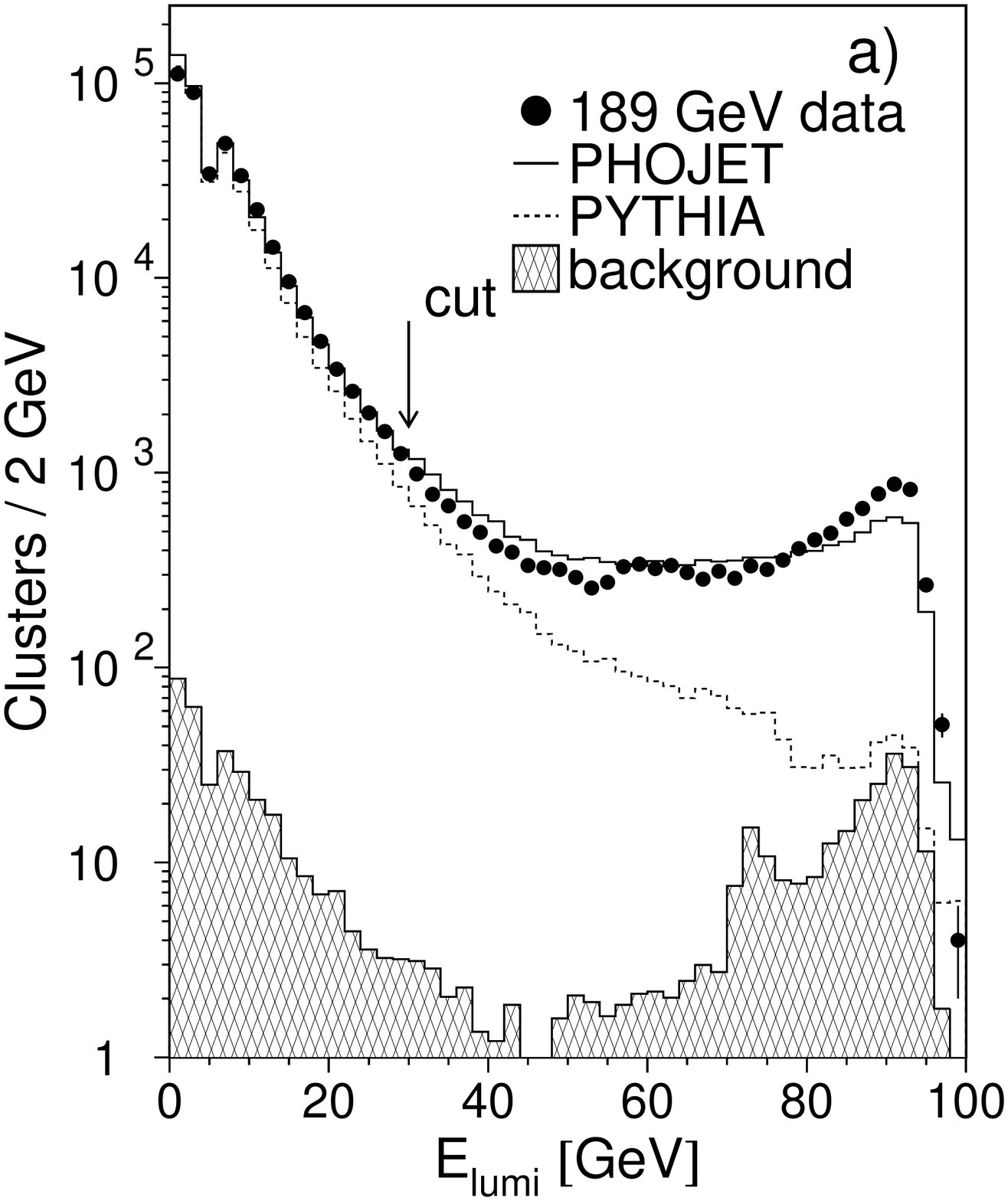}& 
    \includegraphics[width=0.45\textwidth ]{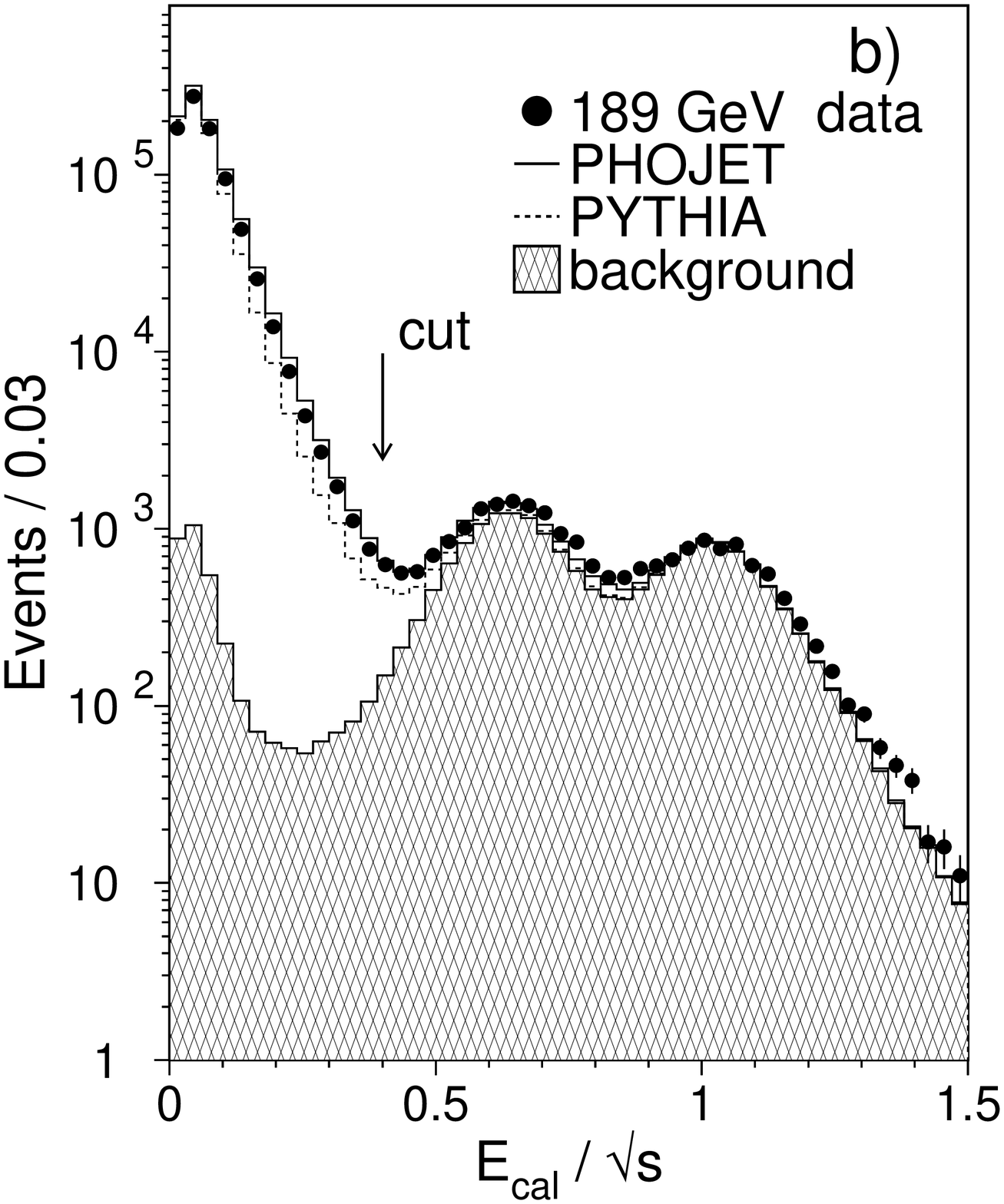}\\
 
 \includegraphics[width=0.45\textwidth ]{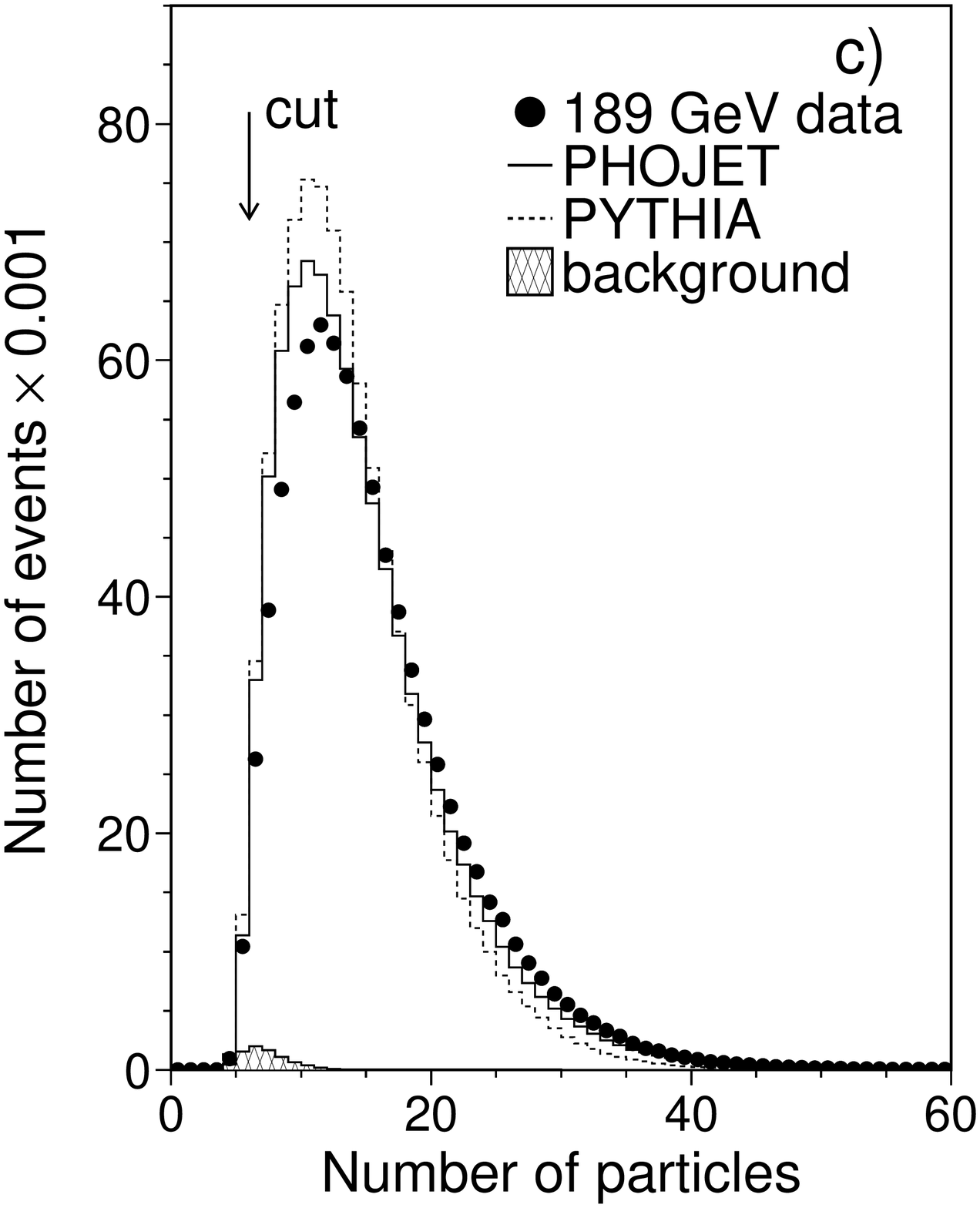}& 
    \includegraphics[width=0.45\textwidth ]{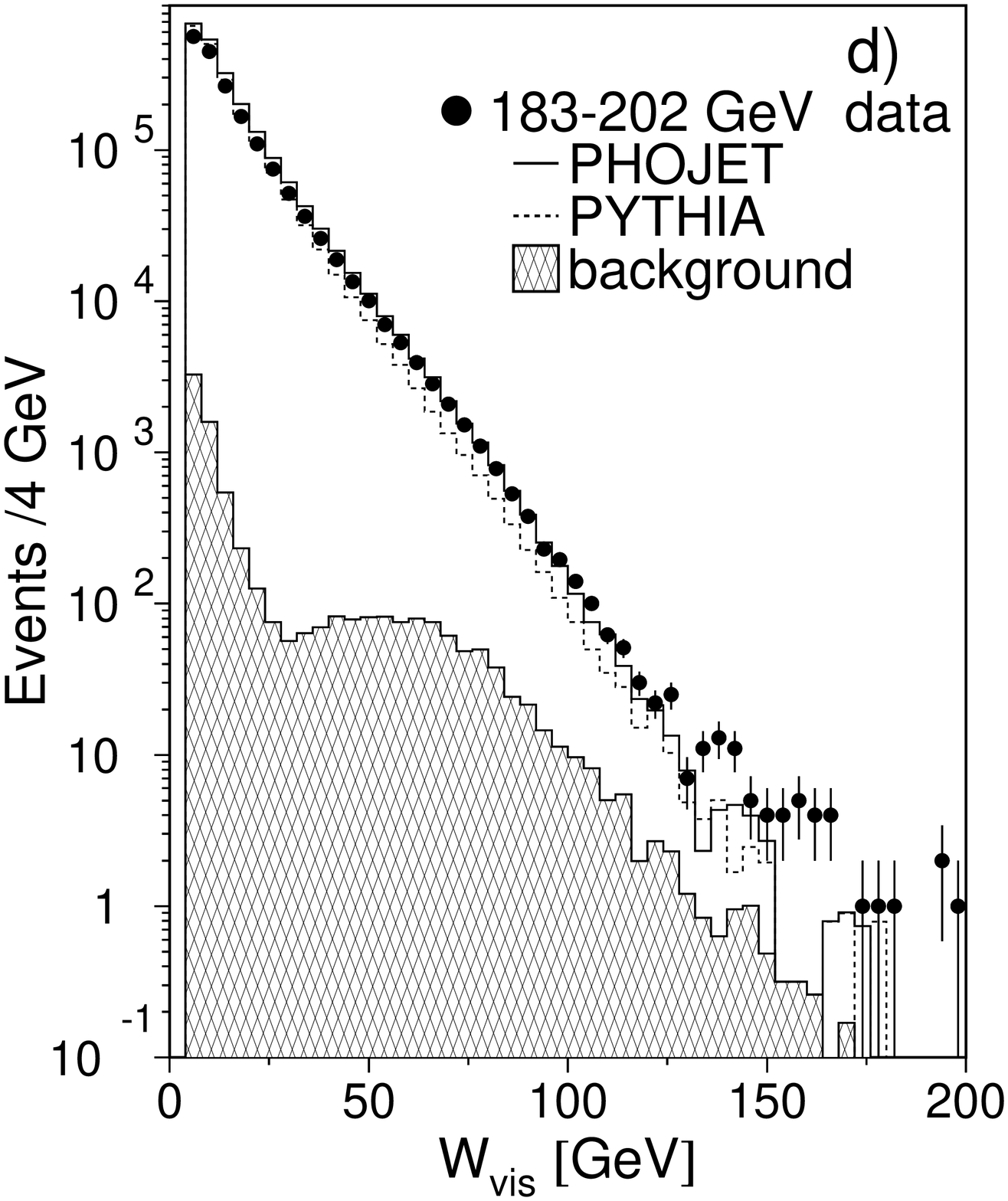}\\
   \end{tabular}           
  \caption{Example of the distributions used for the event
  selection: 
  a) energy in the luminosity  monitor;  
 b) energy $E_{\rm{cal}}$ in the
  electromagnetic and hadronic calorimeters, normalised to  \rts ; 
    c) number of particles measured in the detector. 
   d)  The distribution of the visible  mass \Wvis , for the full data sample
 at $\rts = 183 \GeV - 202 \GeV$.
  The data are  compared  with Monte Carlo predictions. 
  The backgrounds due to \ee annihilation and \eett {}
  are indicated as a shaded area. }
  \label{fig:cal}
 
\end{figure}

Hadronic two-photon
  events are selected by the following criteria :
\begin{itemize}
\item
 To exclude scattered electrons,  events
  with clusters in the luminosity monitor having 
 energy greater than 30 \GeV , 
in a fiducial region of
 33 mrad $\le \theta \le 64$ mrad are rejected.  The 
 virtuality of the interacting photons, $Q^2$, 
  is thus less than 8$\GeV ^2$, with an average
 value  $<Q^2> \sim 1.5 \times  10^{-2} \GeV ^2$.
 The  distribution of low energy clusters  in the  luminosity monitor, presented in  
Figure \ref{fig:cal}a, shows a good agreement with both Monte Carlo programs. 
 When the scattered electron  reaches the detector,
the agreement is maintained with the PHOJET Monte Carlo, while these configurations
are missing in PYTHIA because of a $\rho -$mass cutoff, $Q^{2} \leq {\mathrm m_\rho ^2} $,
applied to the two-photon luminosity
function in the  generation of the events. 

\item 
The total energy in the electromagnetic calorimeter is required to be 
greater than 0.5 \GeV , in order to suppress
 beam-gas and beam-wall backgrounds,
 and less than 50 \GeV , to exclude radiative events, 
 \ee {} \ra {} Z$ \gamma$.
The total  energy deposited in the electromagnetic
and hadronic  calorimeters, $E_{\rm{cal}}$, must be 
less than 40\% of \rts ,
 to exclude  annihilation events, as shown in Figure \ref{fig:cal}b. 

\item 
At  least six particles must be detected, in order to exclude  events containing $\tau$.
A particle is defined \cite{had161} as either a track, a photon 
in the electromagnetic calorimeter, or a
cluster in the hadron calorimeter or 
in the luminosity monitor. 
Clusters in the luminosity monitor 
 are  considered as  pions if their 
 energy is below 5 \GeV ,  as photons otherwise.  
The distribution of the number of particles is presented in Figure \ref{fig:cal}c.
This distribution is not well reproduced by the Monte Carlo simulations.
\end{itemize}
After  selection, the background from beam-gas and beam-wall
interactions is found to be negligible. 
The visible effective mass  of the event, \Wvis , is calculated from the 
four-momenta  of the measured particles.
 The analysis is limited to events with $\Wvise \geq 5 \GeV $.
 \par
Almost 2 million events are selected,
$ 1.6 \times 10^5$ at \rts = 183 \GeV , $ 7.8 \times 10^5$ at \rts = 189 \GeV {}
and $ 1 \times 10^6$ at \rts = 192 $-$ 202 \GeV . The average centre-of-mass energy of this last
sample is  \rts = 198 \GeV .
\par 
 The \Wvis {} spectrum is shown in Figure \ref{fig:cal}d for the total data sample.
The background 
 is below 1\% at low masses, where it is dominated by   two-photon $\tau$-pair production.
 It increases at  high masses,
 due mainly to annihilation processes and  reaches 
 a maximum of  15\%.
 \par
\begin{figure}[htbp]

 \begin{tabular}{cc}

  \includegraphics[width=0.55\textwidth ]{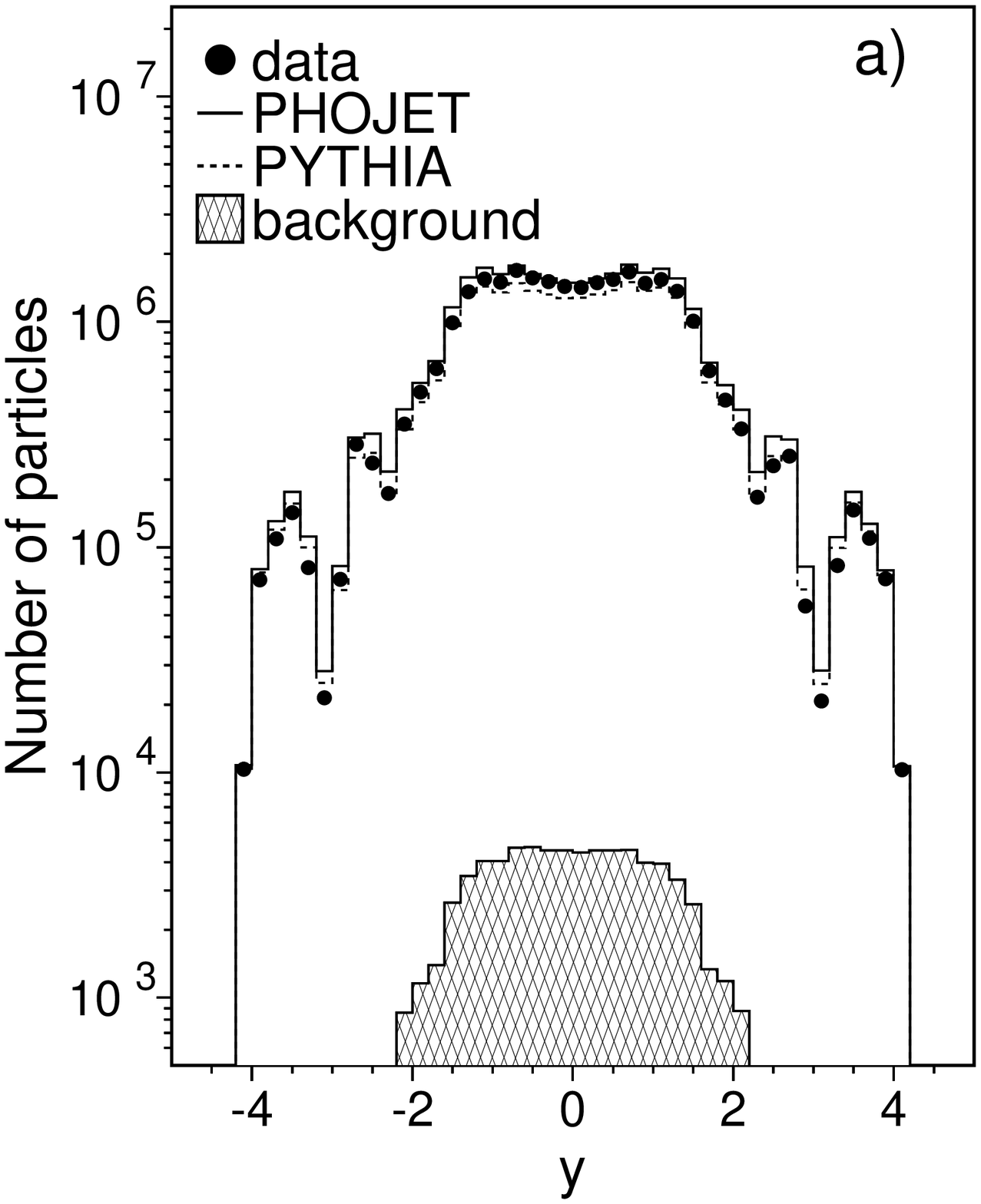}& 
    \includegraphics[width=0.45\textwidth ]{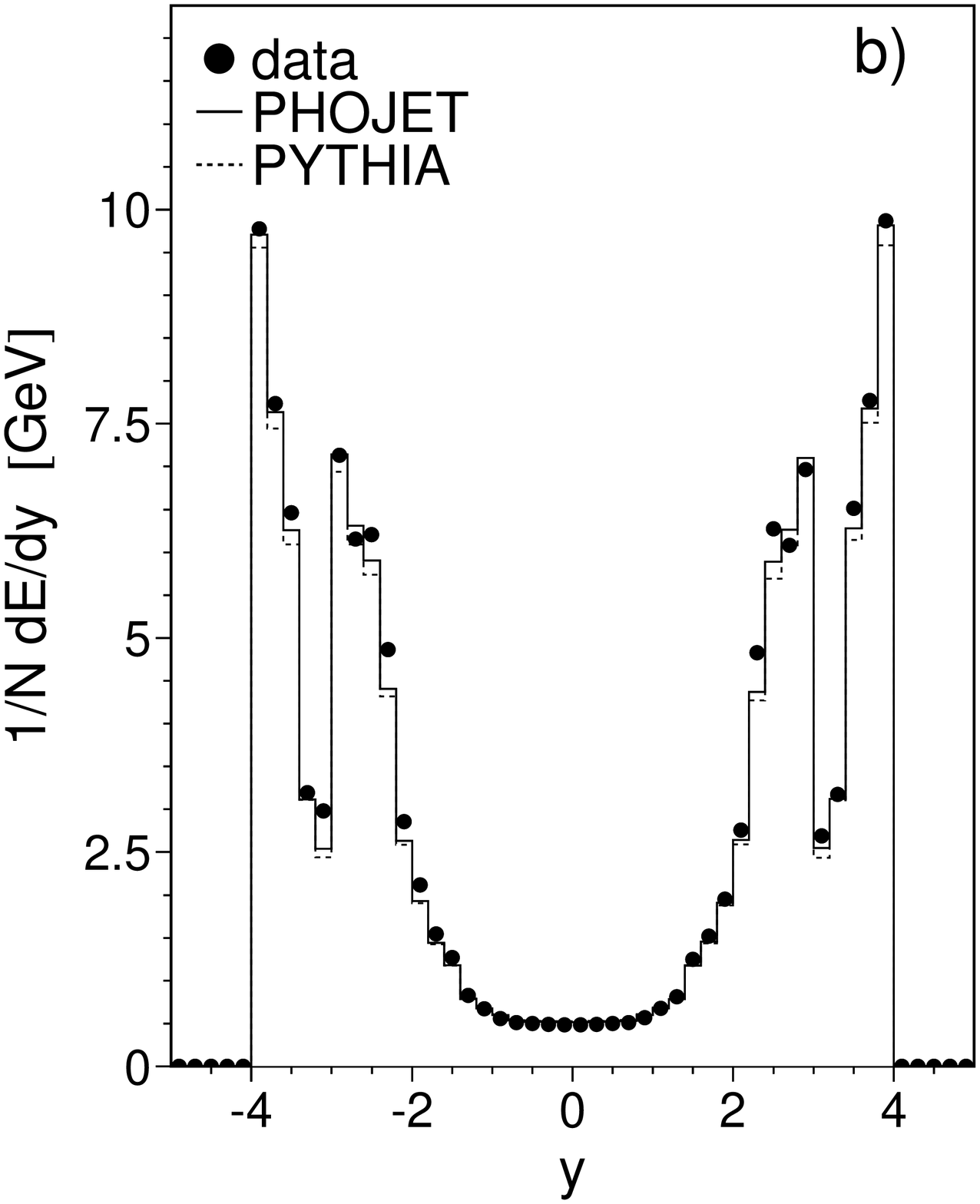}\\
   \end{tabular}           
  \caption{a) The distribution and
  b) the  average energy of the measured particles as a function of the rapidity $y$,
  for the full data sample
 at $\rts =  183 \GeV - 202 \GeV$. 
   The data are  compared  with Monte Carlo predictions. 
  The backgrounds due to \ee annihilation and \eett {}
  are indicated as a shaded area. }
  \label{fig:y}
 
\end{figure}

  The distributions of the rapidity, $y$,  of the particles and of their energy flow
 are compared to the Monte Carlo expectations in Figure \ref{fig:y}. A good agreement is observed
 also in the regions where ${ \mid y \mid} \simeq 3$,  between
 the luminosity monitor and the hadron calorimeter.

\subsection{Unfolding and efficiency}
\begin{figure}

 \begin{tabular}{cc}
  \includegraphics[width=0.48\textwidth ]{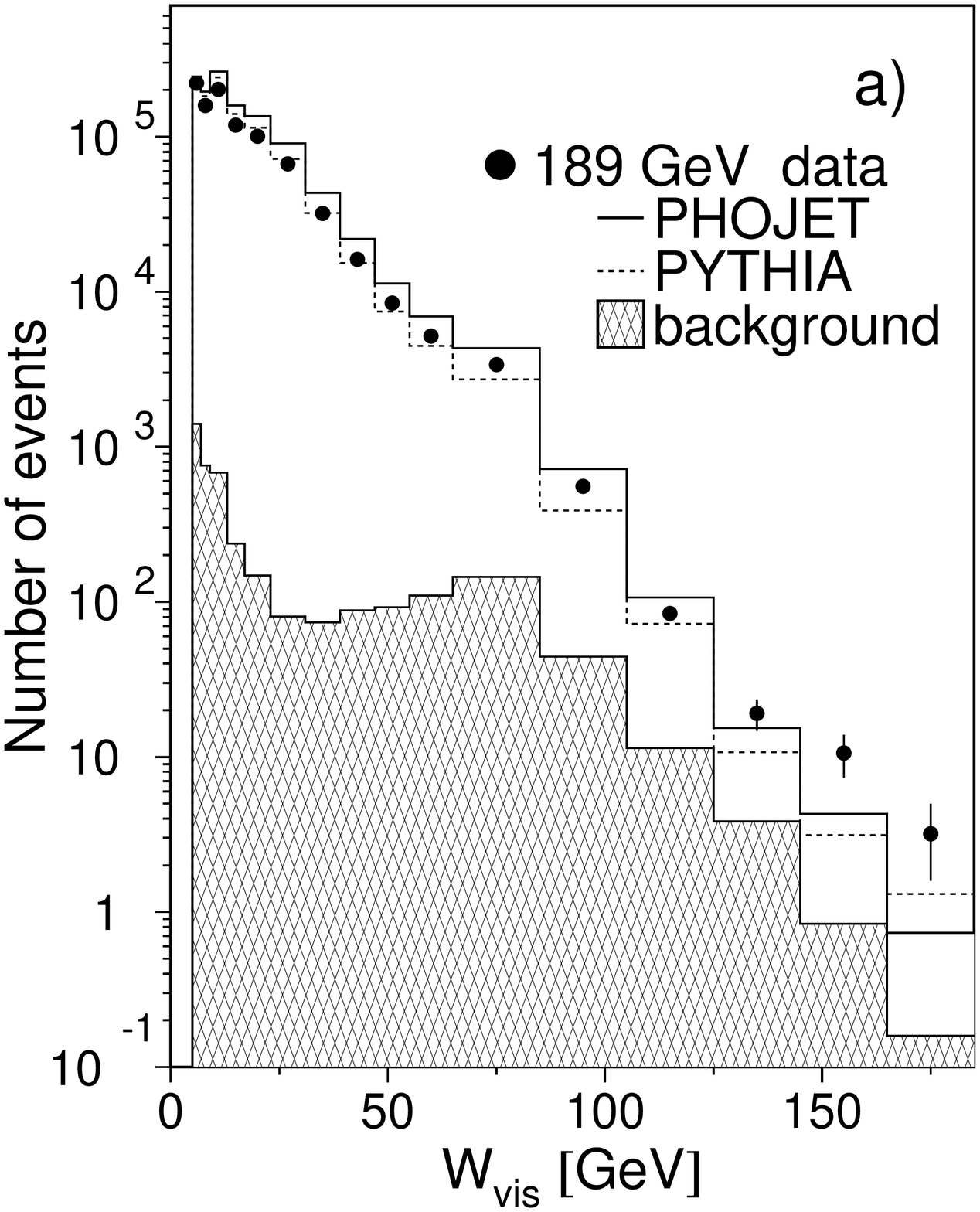}&
  \includegraphics[width=0.48\textwidth]{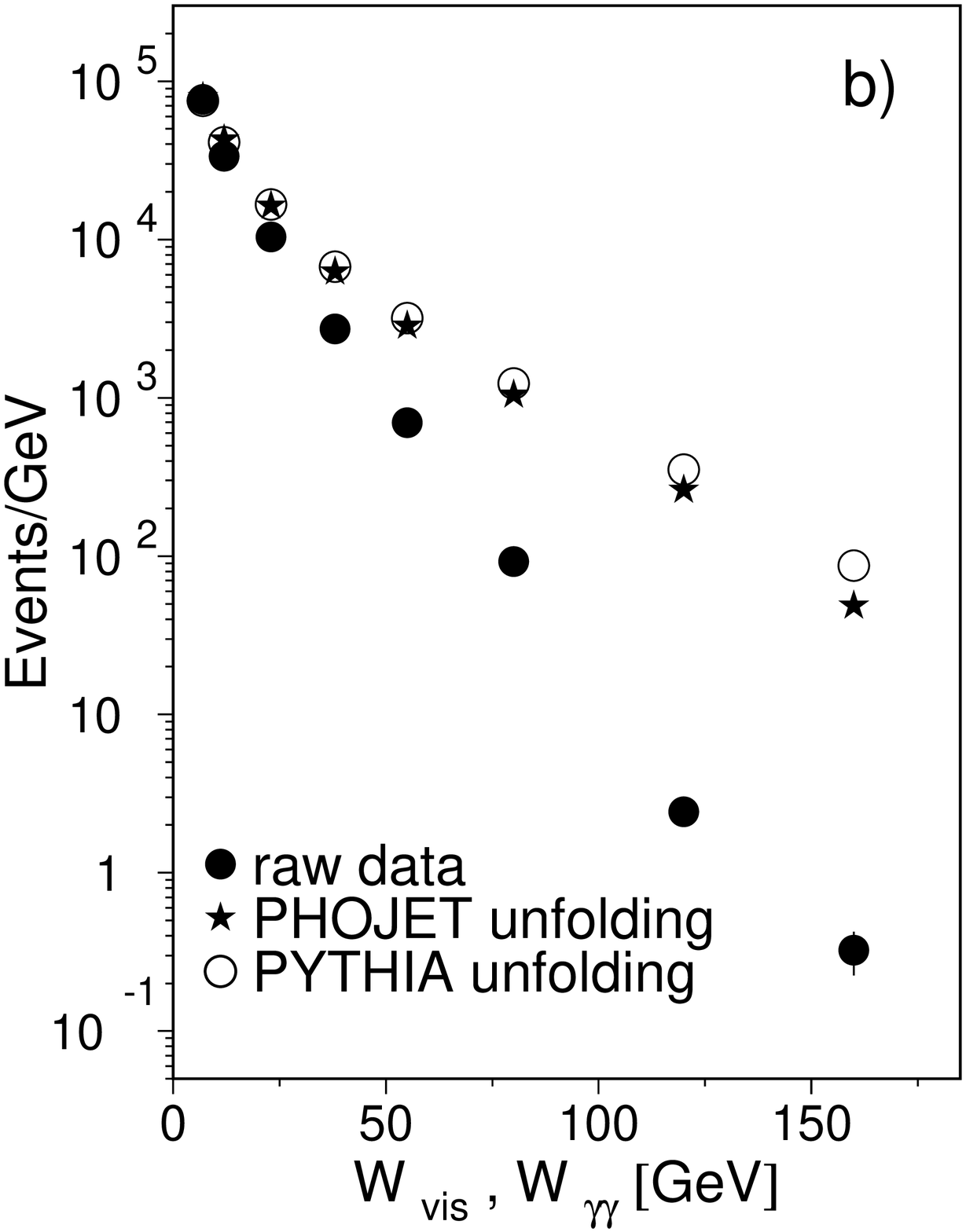}\\
  \end{tabular}
 
 \caption{Example of the unfolding for  data at $\rts =189  \GeV$. 
 a) Distribution of the measured visible mass \Wvis 
 {} in 
 the 16 bins used for the unfolding. 
  b) The measured \Wvis {}   and the 
  resulting \Wgg {} spectra, obtained by unfolding  from \Wvis {} to \Wgg {}
  with the Monte Carlo PHOJET or PYTHIA.} 
  \label{unf}
  
\end{figure}

The distribution of  the two-photon effective mass  \Wgg {}
 is obtained from the visible effective mass \Wvis {}
by the same  unfolding procedure \cite{bayes} used in Reference \citen{had161}. 
For each data sample, the \Wvis {} spectrum is  subdivided in 16 intervals,
 presented in Figure \ref{unf}a,
 and the resulting 
  \Wgg {} distribution in 
 8 intervals, presented in Figure \ref{unf}b. 
The result of the unfolding procedure  depends on the
Monte Carlo used.
Data unfolded with PYTHIA are in general higher than
if unfolded with PHOJET.
After unfolding, the events are corrected for  
 the efficiency,  
using the ratio between selected and 
generated events in each  \Wgg {} interval.
This includes the purely geometrical acceptance  as well as the
efficiencies of the detector and  the 
analysis procedure.
 For \Wgg $> 30$ \GeV {}, the efficiency is rather constant, with a value of about 80\%.
The efficiency obtained with PYTHIA is lower by about 10\%, which may be attributed to   a different modeling of the diffractive
interactions, of difficult detection.

\subsection{Cross  Section Determinations}
\begin{figure}
\begin{center}
  \includegraphics[height=0.6\textheight]{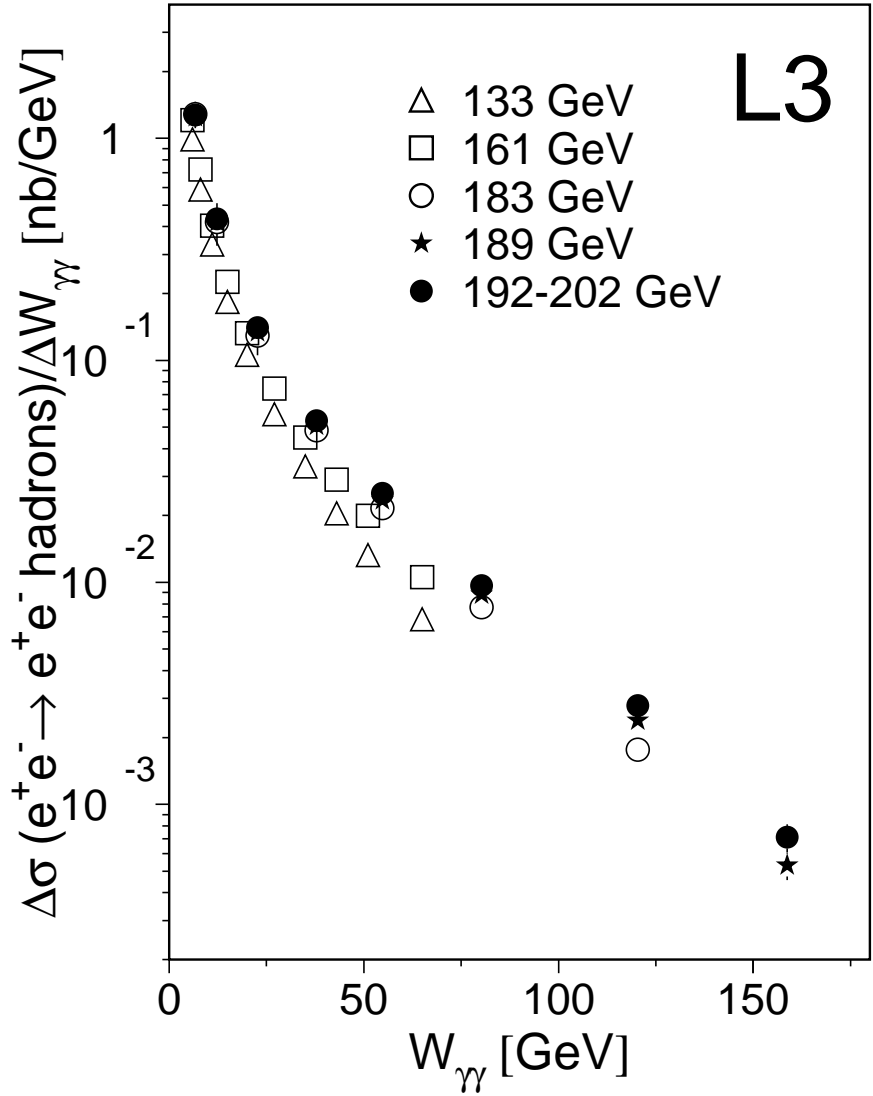} 
\end{center}
 \caption{
 The  cross section \dseeh  {}  measured at 
   $\rts =133 \GeV - 202 \GeV$.  
  Statistical and
   systematic uncertainties are added in quadrature and are often smaller than the symbol size.} 
  \label{crossee}
  
\end{figure}

  The measured cross sections  \seeh {} are given in Table 1 for the three data sets,
  as a function of the \Wgg {} intervals.
  The average of the results obtained by unfolding the data with PYTHIA or PHOJET is used.
  Due to the unfolding procedure,
  the measurements are highly correlated. The correlation matrix, similar for the three data sets,
  is given  in Table 2 for \rts = 189 \GeV .
 The differential cross section $\Delta \sigma /\Delta$\Wgg {} 
is shown in  Figure \ref{crossee},  together with our measurements at 
lower LEP collision energies \cite{had161}.
The fast decrease of the cross section
as a function of \Wgg {} is due to the two photon
luminosity function, ${\cal{L}}_{\gamma \gamma} $,
which depends on $ \Wgge ^{2} /s$. 

\par
 The systematic uncertainties are 
 evaluated for each \Wgg {} bin.
  They are independent of the data sample, 
  inside statistical accuracy.
  They are evaluated as follows and their contribution is listed
  in Table 3.
    \begin{itemize} 
 \item
 Trigger efficiencies: by varying this quantity
 within the accuracy of its determination, of about 10\%.
   \item
  Energy scale of the calorimeters and contribution of the annihilation background:  
  by varying the $ E_{\rm{cal}}$ cut by $\pm$ 10\% of \rts .
  \item
  Uncertainties on the rejection of scattered electrons: by changing the 
  $\rm{E_{lumi}}$ cut from 30 \GeV {} to 50 \GeV. 
  \item 
  Uncertainties on the particle multiplicity:
  by  accepting a minimum number of 
   four or  eight particles instead of six.
  \item
 Uncertainties due to  Monte Carlo statistics are negligible for $ \Wgge <65 \GeV $, but important 
 in the higher  \Wgg {} bins.
  \end{itemize}
 Uncertainties on the energy scale of the small angle calorimeter, 
  evaluated by varying the gain  by a factor two, are negligible.
  The total experimental systematic uncertainty,
   obtained by adding in quadrature all contributions, is also given in Table 3.
The uncertainty related to the Monte Carlo model is given in the last column of Table 3. 
 It is  half of the difference between the results obtained by 
 unfolding the data with PHOJET or PYTHIA and exceeds the experimental 
uncertainty in almost  all bins. 
\par 
To  extract the total cross section of two 
 real photons,  the luminosity function \Lgg 
{}\cite{Budnev} is calculated and 
 the hadronic two-photon process  is extrapolated to  $Q^2  = 0$.
This is done as in Reference \cite{had161} by using an analytical program 
 \cite{Schuler}.
 Depending on the choice of  photon form factors, this calculation  varies
  of $\pm$5\%.
\par
The cross sections obtained for the three data sets  are compatible 
within the experimental uncertainties and are presented in Table 4. 
As expected from the study of the experimental systematics, the largest differences are observed in the first and last bins.
 The combined value is 
also given
in Table 4 and in Figure \ref{fig:fit}a 
 with the statistical uncertainties obtained from the unfolding and the
experimental systematics. The values obtained by unfolding the data with 
the two Monte Carlo programs separately are shown in Figure \ref{fig:fit}b and 
can be obtained  from the last column of Table 4.

\section{Comparison with Theoretical Models}
\subsection{Regge parametrisation}
 The total  cross sections for hadron-hadron,
$\sigma _{\mathrm{p} \mathrm{p}}$, photon-hadron, $ \sigma _{\gamma \mathrm{p}}$,
and photon-photon, $ \sigma _{\gamma \gamma}$,
 production of hadrons
  show a characteristic steep decrease in the region of 
 low centre-of-mass energy, followed by  a slow rise at high energies.
From Regge  theory \cite{Collins} this behaviour is understood as the 
consequence of the  exchange of Regge trajectories, $\alpha (t)$,
in the $t$-channel. The total cross section takes the form
$\sigma_{\mathrm{tot}} \propto s^{ \alpha  (0) -1}$. 
The  low energy region
is sensitive to the
exchange of a Reggeon $ R$ ($R = \rho$, $\omega $, f, a ..), with $\alpha_{R}(0)\simeq 0.5$. 
 At high energies, the  Pomeron exchange dominates, with $\alpha _{P} (0) \simeq 1$.
  A parametrisation   
of the form 
\begin{equation} \sigma_{\mathrm{tot}} 
= A \, s^{\epsilon} \, + \, B \, s^{-\eta}
\end{equation}
accounts for the energy behaviour of all hadronic
and photoproduction total cross sections,
the powers of $s$ being universal\cite{DL}.
This is confirmed by the recent compilation of the
total cross section data \cite{PDG}
 where a fit of Equation 1 for all hadron 
  total cross
 sections  gives a result  compatible with the universal  
 values 
$\epsilon =0.093 \pm 0.002$ and $\eta =0.358 \pm 0.015$.
 The coefficients $A$ and $B$ are process
and $Q^2$ dependent.
If photons behave predominantly like hadrons, this expression may  also be valid for the two-photon
total hadronic cross section, with $s =\Wgge ^2$.
\par
\begin{figure}
  \begin{tabular}{ll}
  \includegraphics[width=0.45\textwidth]{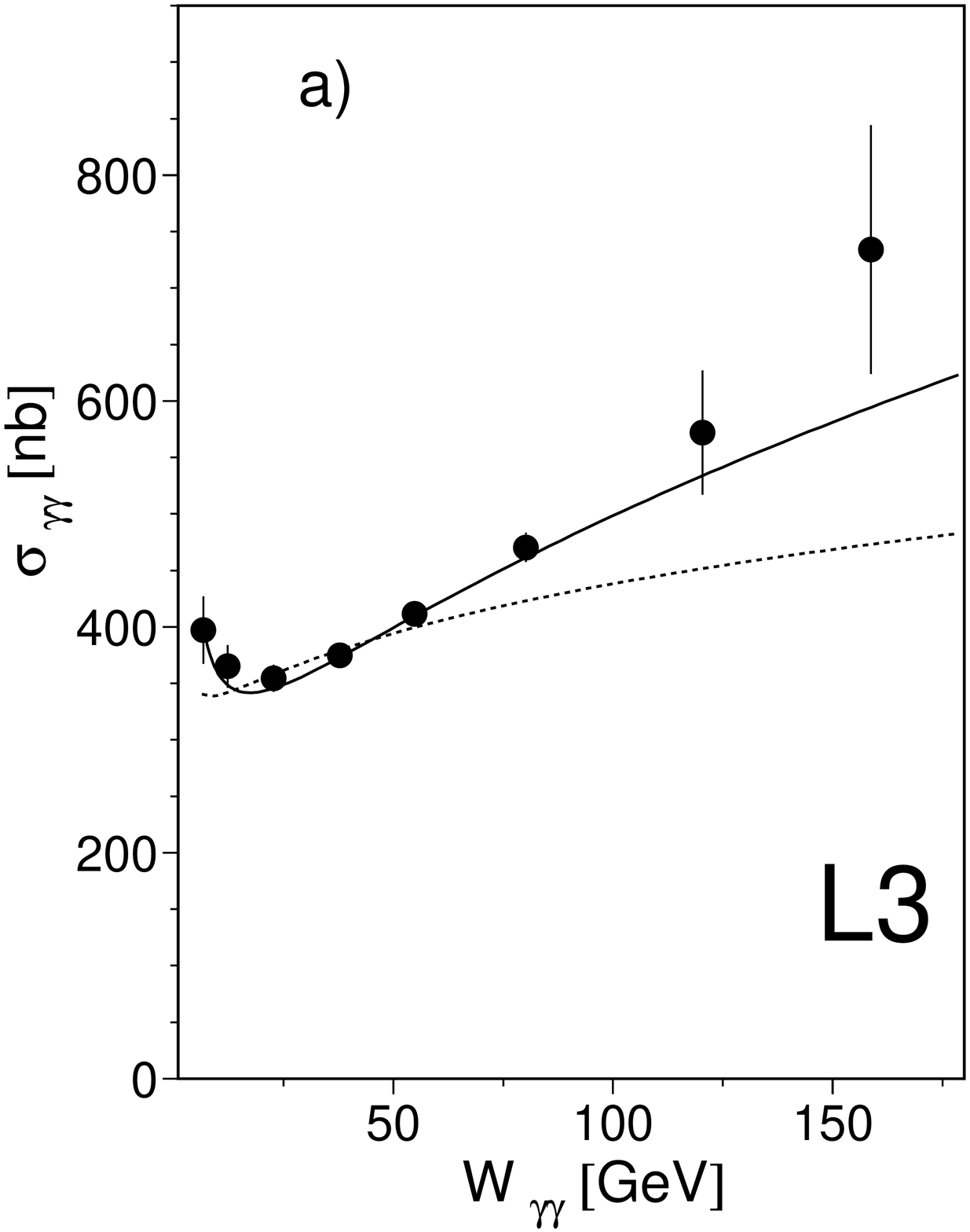}& 
  \includegraphics[width=0.45\textwidth]{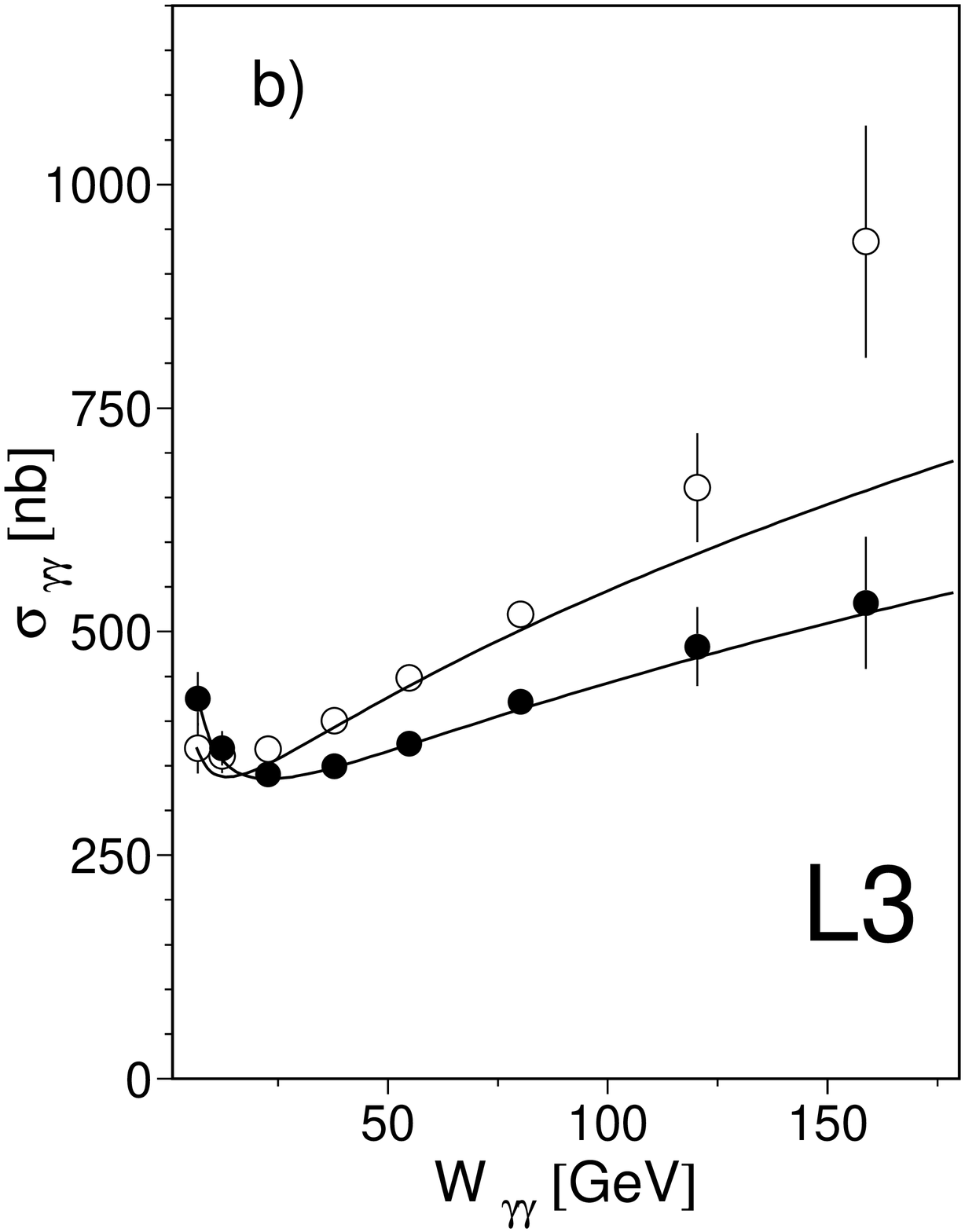} \\
\end{tabular} 
  \caption{ 
    The two-photon total cross section from the combined data at  $\rts = 183 \GeV -  202  \GeV $.
  a) The  average result, obtained by unfolding the data with the two Monte Carlo 
  models, is used.
   Two Regge fits, described in the text, are superimposed
   to the data. The continous line corresponds to  the fit with the coefficient $\epsilon$
   left as a free parameter, the dashed line is the fit with $\epsilon$ fixed to 0.093.     
   b) The two-photon total cross section obtained by correcting the 
    data sample with PHOJET
(full points) and with PYTHIA (open points). The Regge fits of Table 5  are superimposed
 to the data. The   statistical and experimental
   systematic uncertainties are added in quadrature. } 
 
  \label{fig:fit}
\end{figure}

 
Considering only the experimental uncertainties, statistical and  systematic,
several Regge fits are performed on   the data and
their results  are presented in  Table 5.
The exponent $\eta$ is fixed to the universal value, since the low mass range is too small
to be sensitive to this parameter.
When the \Wgg {} interval is restricted to $5\GeV - 65 \GeV $, a range similar to the one 
covered by our previous data \cite{had161}, similar values of the parameters $A$ and $B$ are 
obtained. In this limited interval the data are compatible with the universal value of $\epsilon$.
Extending the range to the whole \Wgg {} interval, the fit 
  with the exponents 
$\epsilon$ and $\eta$  fixed to the universal value, dashed line  in Figure \ref{fig:fit}a, 
 does not
represent  the $\sigma _{\gamma \gamma}$ energy dependence.
A fit with $A$, $B$ and  $\epsilon $ as free parameters, represented as a  full line
in Figure \ref{fig:fit}a,
 gives 
$\epsilon = 0.225 \pm 0.021$ with a confidence level of   4\%.  
This value is  more than a factor two higher than the universal value. It is  independent
of the Monte Carlo model used to correct the data,
as shown in Table 5  and in Figure \ref{fig:fit}b.
\par
 The fitted value of $\epsilon$ is strongly correlated to the
Reggeon component. To  avoid this correlation, we fit only the Pomeron exchange for sufficiently
high  \Wgg {} values. The  results, using a different initial value of \Wgg ,
 are listed in the second part of Table 5 .
The value of  $\epsilon$ increases by increasing the lower 
mass cutoff, thus indicating  that its value is not universal, but it reflects the 
 onset of QCD phenomena, as $\epsilon$ increases with increasing \Wgg .

\subsection{Models for {\boldmath \gg } total cross sections}
\begin{figure}
  
  \begin{tabular}{ll}
   \includegraphics[height=0.4\textheight]{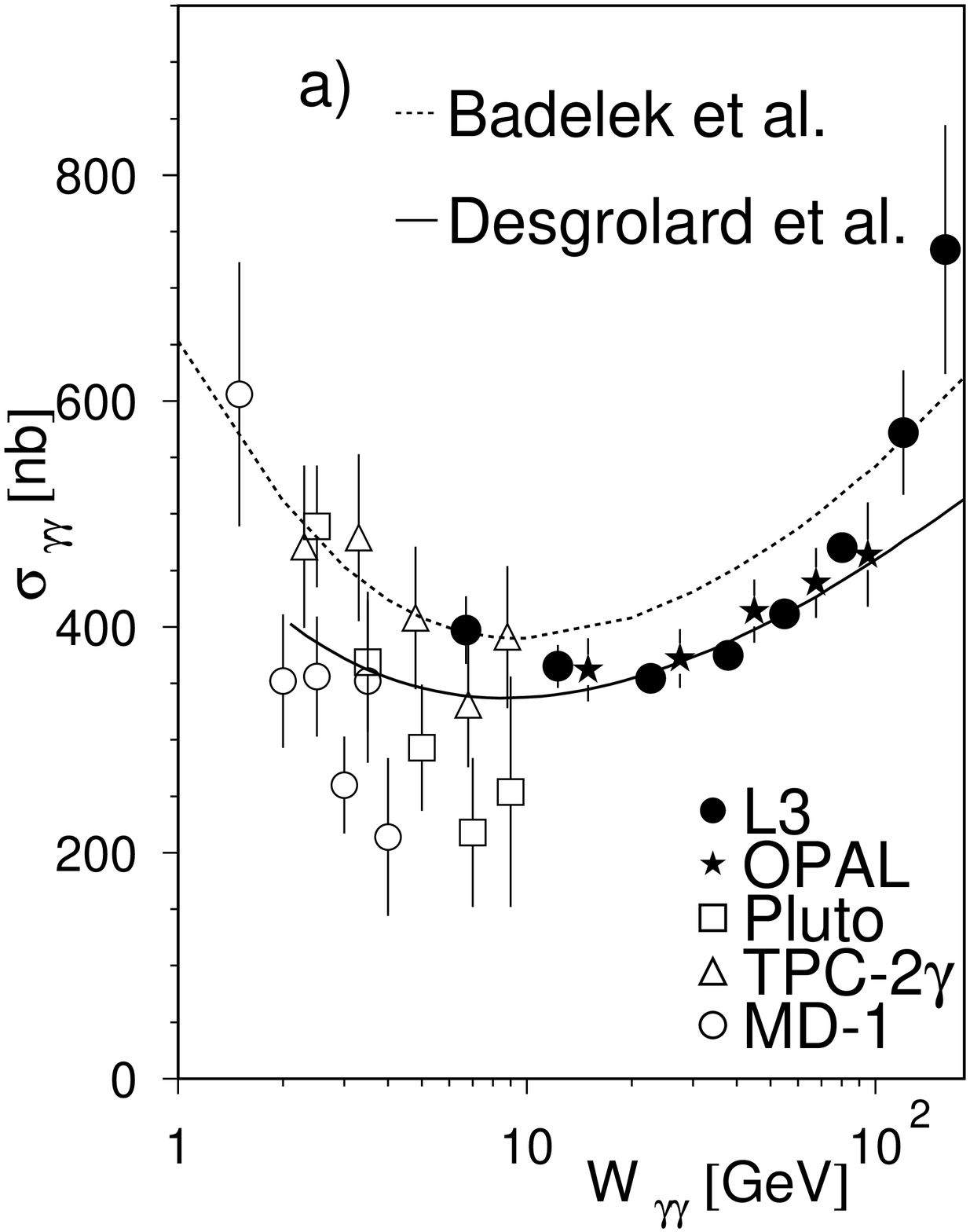} 
  \includegraphics[height=0.4\textheight]{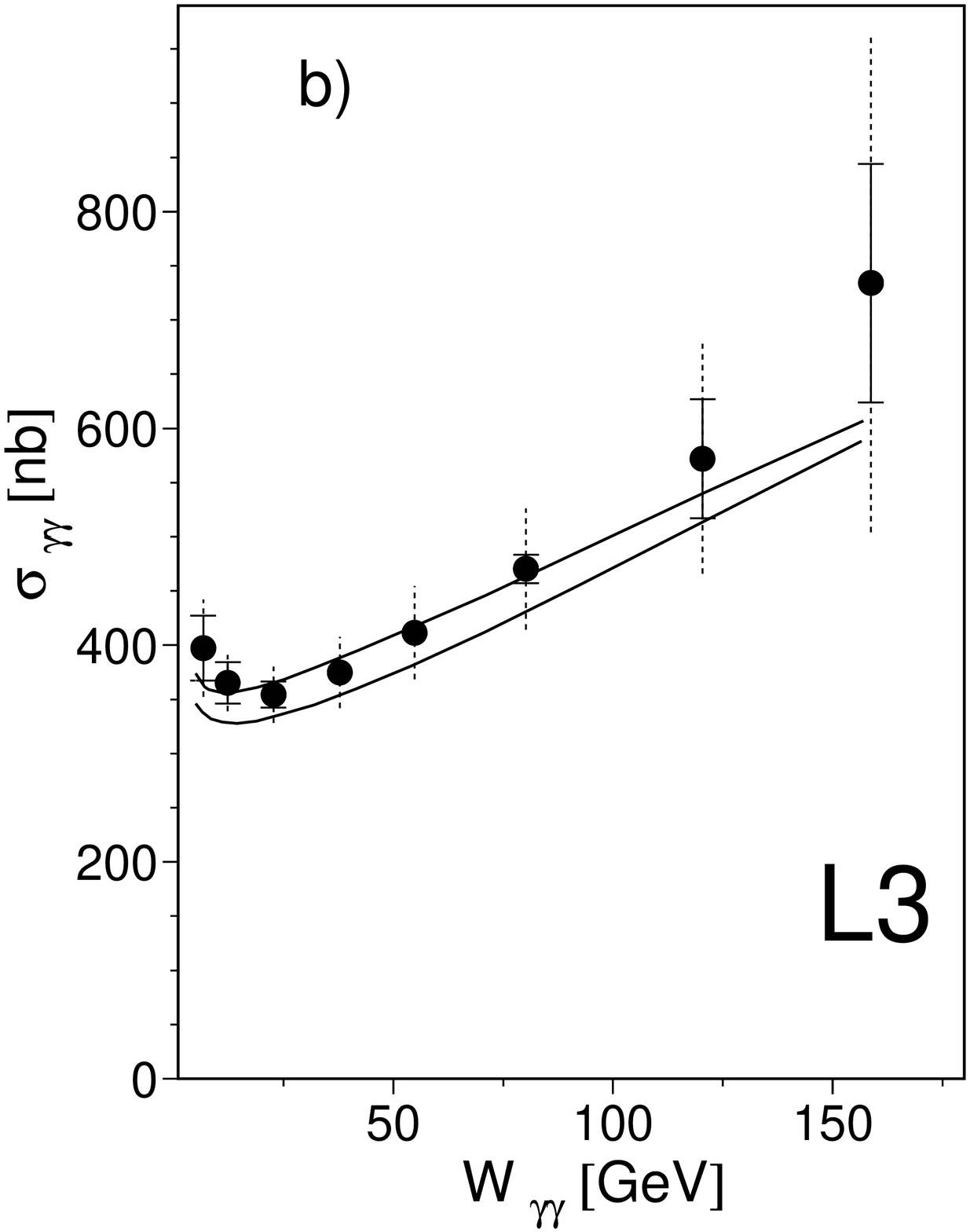} 
 \end{tabular} 
  \caption{ The two-photon total cross sections 
compared to various models.
   a) The predictions of References \protect\citen{badelek} 
    and  \protect\citen{martynov} are compared to all
    two-photon total cross section data \protect\cite{opal ,old}.
    b) Predictions of the minijet model \protect\cite{giulia}; the two lines correspond to
    different choices of the model parameters.
   The statistical and experimental systematic uncertainties are added in quadrature.
    The  uncertainties due to Monte Carlo models and to
    the two-photon luminosity function are included in the  dashed lines in b).} 
  \label{fig:model}
\end{figure}

Several models \cite{SS,badelek,martynov} were recently compared  to the L3 and OPAL \cite{opal}
measurements.  Their predictions for the two-photon total cross section 
are typically derived 
 from measurements of 
proton-proton and photoproduction total cross sections
via the  factorization relation:
$\sigma _{\gamma \gamma} \approx  \sigma _{\gamma \mathrm{p}}^2 /
\sigma _{\mathrm{p} \mathrm{p}}$ \cite{oldR}.
In general, these models give an energy dependence of the cross section similar to the
universal fit discussed above.
 Two examples \cite{badelek , martynov} 
 are shown in Figure \ref{fig:model}a in comparison with  the results of previous experiments \cite{old},
 those presented in this letter and those of OPAL. While the measurements at the low energy colliders
 present a wide spread, a good agreement is found between L3 and  OPAL in the common \Wgg {} range, $10 \GeV \le \Wgge \le 110$
 \GeV .
 Good agreement is also found if the data, unfolded separately with either PHOJET or PYTHIA,  are compared.
 In this \Wgg {} region, a model \cite{martynov} reproduces well the data and the 
 predictions of  the other \cite{badelek} are too high by 20\%.
 However, for both lower and higher values of \Wgg , the L3 data show a much steeper energy dependence than the
 theoretical predictions.

\par 
In the Regge theory, the Pomeron intercept is 1, yielding a  constant total
cross section. When the rise of the proton-proton 
 total cross section was first observed, it was explained \cite{rise}
with an increase of the number of hard partonic interactions.
The predictions of  a model \cite{giulia} that calculates such effects,
using an eikonalized prescription to enforce unitarity, are  shown in Figure \ref{fig:model}b.
The parameters of the model are determined from 
 photoproduction data and  the L3 results are well inside the uncertainty 
related to this extrapolation.  
\par 
Models with two Pomerons were recently  proposed \cite{dipole}
 to explain the fast energy increase of charm production at HERA.
 In this model,  the  `soft' and the  `hard' Pomeron have different 
 intercepts.
 Because of the \qqbar {} component in the photon wave-function,
 the `hard' Pomeron can contribute to the two-photon cross section even  at $Q^2=0$.
 Thus a more rapid energy dependence for $\sigma _{\gamma \gamma}$ is expected.
  The  increase in $\epsilon$ 
  with larger values of \Wgg , as listed in the
second part of Table 5, is consistent with such 
 a contribution of  the `hard' Pomeron.

\section*{Acknowledgements}

We
express our gratitude to the CERN accelerator divisions for
the excellent performance of the LEP machine.
We acknowledge with appreciation the effort of  engineers,
technicians and support staff who have participated in the
construction and maintenance of this experiment.

%
%%%%%%%%%%%%%%%%%%%%%%%%%%%%%%%%%%%%%%%%%%%%%%%%%%%%%%%%%%%%%%%%%%%%%%%%%%%%%%
%
%%%%%%%%%%%%%%%%%%%%%%%%%%%%%%%%%%%%%%%%%%%%%%%%%%%%%%%%%%%%%%%%%%%%%%%%%%%%%%%
% The author list
%%%%%%%%%%%%%%%%%%%%%%%%%%%%%%%%%%%%%%%%%%%%%%%%%%%%%%%%%%%%%%%%%%%%%%%%%%%%%%%
%
\newpage
\section*{Author List}
\typeout{   }     
\typeout{Using author list for paper 235 -- ? }
\typeout{$Modified: Fri Jan 26 2001 by smele $}
\typeout{!!!!  This should only be used with document option a4p!!!!}
\typeout{   }
%
%
%
%  L A T E X  version!!
%
%
% Make sure that the Lep package has been used!
%\input{Lep.sty}%
%
%\ifx\LepCalled\undefined%
%\typeout{     }%
%\typeout{!!!!!!!!!!!!!!!!!!!!!!!!!!!!!!!!!!!!!!!!!!!!!!!!!!!!!!!!!!!}%
%\typeout{Yikes.  You haven't used the Lep package!}%
%\typeout{Please put \protect\usepackage\protect{Lep\protect} in your preamble,
%         followed by}%
%\typeout{\protect\Lep\protect{1\protect} or \protect\Lep\protect{2\protect}}%
%\typeout{     }%
%\typeout{For now you will get a Lep phase 2 authorlist (may not be right!).}%
%\typeout{!!!!!!!!!!!!!!!!!!!!!!!!!!!!!!!!!!!!!!!!!!!!!!!!!!!!!!!!!!!}%
%\typeout{     }%
%\Lep{2}\fi%

\newcount\tutecount  \tutecount=0
\def\tutenum#1{\global\advance\tutecount by 1 \xdef#1{\the\tutecount}}
\def\tute#1{$^{#1}$}
\tutenum\aachen            % 1
\tutenum\nikhef            % 2
\tutenum\mich              % 3
\tutenum\lapp              % 4
\tutenum\basel             % 5
\tutenum\lsu               % 6
\tutenum\beijing           % 7
\tutenum\berlin            % 8
\tutenum\bologna           % 9 
\tutenum\tata              % 10
\tutenum\ne                % 11
\tutenum\bucharest         % 12
\tutenum\budapest          % 13
\tutenum\mit               % 14 
\tutenum\debrecen          % 15
\tutenum\florence          % 16
\tutenum\cern              % 17 
\tutenum\wl                % 18 
\tutenum\geneva            % 19
\tutenum\hefei             % 20
\tutenum\lausanne          % 21
\tutenum\lecce             % 22
\tutenum\lyon              % 23
\tutenum\madrid            % 24
\tutenum\milan             % 25
\tutenum\moscow            % 26
\tutenum\naples            % 27
\tutenum\cyprus            % 29
\tutenum\nymegen           % 30
\tutenum\caltech           % 31
\tutenum\perugia           % 32
\tutenum\peters            % 33
\tutenum\cmu               % 34
\tutenum\potenza           % 35
\tutenum\prince            % 36
\tutenum\riverside         % 37
\tutenum\rome              % 38
\tutenum\salerno           % 39
\tutenum\ucsd              % 40
\tutenum\sofia             % 41
\tutenum\korea             % 42
\tutenum\alabama           % 43
\tutenum\utrecht           % 44
\tutenum\purdue            % 45
\tutenum\psinst            % 46
\tutenum\zeuthen           % 47
\tutenum\eth               % 48
\tutenum\hamburg           % 49
\tutenum\taiwan            % 50
\tutenum\tsinghua          % 51

{
\parskip=0pt
\noindent
{\bf The L3 Collaboration:}
\ifx\selectfont\undefined%  old style font selection
 \baselineskip=10.8pt
 \baselineskip\baselinestretch\baselineskip
 \normalbaselineskip\baselineskip
 \ixpt
\else%                      new style font selection
 \fontsize{9}{10.8pt}\selectfont
\fi
\medskip
\tolerance=10000
\hbadness=5000
\raggedright
\hsize=162truemm\hoffset=0mm
\def\r{\rlap,}
\noindent

M.Acciarri\r\tute\milan\
P.Achard\r\tute\geneva\ 
O.Adriani\r\tute{\florence}\ 
M.Aguilar-Benitez\r\tute\madrid\ 
J.Alcaraz\r\tute\madrid\ 
G.Alemanni\r\tute\lausanne\
J.Allaby\r\tute\cern\
A.Aloisio\r\tute\naples\ 
M.G.Alviggi\r\tute\naples\
G.Ambrosi\r\tute\geneva\
% ...
\newpage
%\rule{\textwidth}{0.4pt}
\begin{list}{A}{\itemsep=0pt plus 0pt minus 0pt\parsep=0pt plus 0pt minus 0pt
                \topsep=0pt plus 0pt minus 0pt}
\item[\aachen]
 I. Physikalisches Institut, RWTH, D-52056 Aachen, FRG$^{\S}$\\
 III. Physikalisches Institut, RWTH, D-52056 Aachen, FRG$^{\S}$
\item[\nikhef] National Institute for High Energy Physics, NIKHEF, 
     and University of Amsterdam, NL-1009 DB Amsterdam, The Netherlands
\item[\mich] University of Michigan, Ann Arbor, MI 48109, USA
\item[\lapp] Laboratoire d'Annecy-le-Vieux de Physique des Particules, 
     LAPP,IN2P3-CNRS, BP 110, F-74941 Annecy-le-Vieux CEDEX, France
\item[\basel] Institute of Physics, University of Basel, CH-4056 Basel,
     Switzerland
\item[\lsu] Louisiana State University, Baton Rouge, LA 70803, USA
\item[\beijing] Institute of High Energy Physics, IHEP, 
  100039 Beijing, China$^{\triangle}$ 
\item[\berlin] Humboldt University, D-10099 Berlin, FRG$^{\S}$
\item[\bologna] University of Bologna and INFN-Sezione di Bologna, 
     I-40126 Bologna, Italy
\item[\tata] Tata Institute of Fundamental Research, Bombay 400 005, India
\item[\ne] Northeastern University, Boston, MA 02115, USA
\item[\bucharest] Institute of Atomic Physics and University of Bucharest,
     R-76900 Bucharest, Romania
\item[\budapest] Central Research Institute for Physics of the 
     Hungarian Academy of Sciences, H-1525 Budapest 114, Hungary$^{\ddag}$
\item[\mit] Massachusetts Institute of Technology, Cambridge, MA 02139, USA
\item[\debrecen] KLTE-ATOMKI, H-4010 Debrecen, Hungary$^\P$
\item[\florence] INFN Sezione di Firenze and University of Florence, 
     I-50125 Florence, Italy
\item[\cern] European Laboratory for Particle Physics, CERN, 
     CH-1211 Geneva 23, Switzerland
\item[\wl] World Laboratory, FBLJA  Project, CH-1211 Geneva 23, Switzerland
\item[\geneva] University of Geneva, CH-1211 Geneva 4, Switzerland
\item[\hefei] Chinese University of Science and Technology, USTC,
      Hefei, Anhui 230 029, China$^{\triangle}$
\item[\lausanne] University of Lausanne, CH-1015 Lausanne, Switzerland
\item[\lecce] INFN-Sezione di Lecce and Universit\`a Degli Studi di Lecce,
     I-73100 Lecce, Italy
\item[\lyon] Institut de Physique Nucl\'eaire de Lyon, 
     IN2P3-CNRS,Universit\'e Claude Bernard, 
     F-69622 Villeurbanne, France
\item[\madrid] Centro de Investigaciones Energ{\'e}ticas, 
     Medioambientales y Tecnolog{\'\i}cas, CIEMAT, E-28040 Madrid,
     Spain${\flat}$ 
\item[\milan] INFN-Sezione di Milano, I-20133 Milan, Italy
\item[\moscow] Institute of Theoretical and Experimental Physics, ITEP, 
     Moscow, Russia
\item[\naples] INFN-Sezione di Napoli and University of Naples, 
     I-80125 Naples, Italy
\item[\cyprus] Department of Natural Sciences, University of Cyprus,
     Nicosia, Cyprus
\item[\nymegen] University of Nijmegen and NIKHEF, 
     NL-6525 ED Nijmegen, The Netherlands
\item[\caltech] California Institute of Technology, Pasadena, CA 91125, USA
\item[\perugia] INFN-Sezione di Perugia and Universit\`a Degli 
     Studi di Perugia, I-06100 Perugia, Italy   
\item[\peters] Nuclear Physics Institute, St. Petersburg, Russia
\item[\cmu] Carnegie Mellon University, Pittsburgh, PA 15213, USA
\item[\potenza] INFN-Sezione di Napoli and University of Potenza, 
     I-85100 Potenza, Italy
\item[\prince] Princeton University, Princeton, NJ 08544, USA
\item[\riverside] University of Californa, Riverside, CA 92521, USA
\item[\rome] INFN-Sezione di Roma and University of Rome, ``La Sapienza",
     I-00185 Rome, Italy
\item[\salerno] University and INFN, Salerno, I-84100 Salerno, Italy
\item[\ucsd] University of California, San Diego, CA 92093, USA
\item[\sofia] Bulgarian Academy of Sciences, Central Lab.~of 
     Mechatronics and Instrumentation, BU-1113 Sofia, Bulgaria
\item[\korea]  Laboratory of High Energy Physics, 
     Kyungpook National University, 702-701 Taegu, Republic of Korea
\item[\alabama] University of Alabama, Tuscaloosa, AL 35486, USA
\item[\utrecht] Utrecht University and NIKHEF, NL-3584 CB Utrecht, 
     The Netherlands
\item[\purdue] Purdue University, West Lafayette, IN 47907, USA
\item[\psinst] Paul Scherrer Institut, PSI, CH-5232 Villigen, Switzerland
\item[\zeuthen] DESY, D-15738 Zeuthen, 
     FRG
\item[\eth] Eidgen\"ossische Technische Hochschule, ETH Z\"urich,
     CH-8093 Z\"urich, Switzerland
\item[\hamburg] University of Hamburg, D-22761 Hamburg, FRG
\item[\taiwan] National Central University, Chung-Li, Taiwan, China
\item[\tsinghua] Department of Physics, National Tsing Hua University,
      Taiwan, China
\item[\S]  Supported by the German Bundesministerium 
        f\"ur Bildung, Wissenschaft, Forschung und Technologie
\item[\ddag] Supported by the Hungarian OTKA fund under contract
numbers T019181, F023259 and T024011.
\item[\P] Also supported by the Hungarian OTKA fund under contract
  numbers T22238 and T026178.
\item[$\flat$] Supported also by the Comisi\'on Interministerial de Ciencia y 
        Tecnolog{\'\i}a.
\item[$\sharp$] Also supported by CONICET and Universidad Nacional de La Plata,
        CC 67, 1900 La Plata, Argentina.
\item[$\diamondsuit$] Also supported by Panjab University, Chandigarh-160014, 
        India.
\item[$\triangle$] Supported by the National Natural Science
  Foundation of China.
\end{list}
}
\vfill

%%% Local Variables: 
%%% mode: latex
%%% TeX-master: t
%%% End:

\newpage
%

%%%%%%%%%%%%%%%%%%%%%%%%%%%%%%%%%%%%%%%%%%%%%%%%%%%%%%%%%%%%%%%%%%%%
% figures
%%%%%%%%%%%%%%%%%%%%%%%%%%%%%%%%%%%%%%%%%%%%%%%%%%%%%%%%%%%%%%%%%%%%

\newpage
\begin{table}
    \label{tab:decross}
  \begin{center} 
\begin{tabular}{|c|c|c|c|c|}
  \hline
  & \rts  &183 \GeV    & 189 \GeV   & $198$ \GeV \\  
 
 \hline
 $\Delta$ \Wgg (\GeV )  & $<\Wgge >$(\GeV )  & \see (nb)  & \see (nb)   & \see (nb)\\                             
  \hline
$\phantom{0}\phantom{0}5-9\phantom{0}\phantom{0}$ &\phantom{0}\phantom{0}6.7&5.145 $\pm$ 0.025 &4.996 $\pm$ 0.009&5.093 $\pm$ 0.008 \\

$\phantom{0}\phantom{0}9-17\phantom{0}$ &\phantom{0}12.3&3.358 $\pm$ 0.013 &3.350 $\pm$ 0.006&3.466 $\pm$ 0.005\\

$\phantom{0}17-31\phantom{0}$&\phantom{0}22.7&1.812 $\pm$ 0.007 &1.880 $\pm$ 0.004&1.962 $\pm$ 0.003 \\

$\phantom{0}31-47\phantom{0}$&\phantom{0}37.8&0.776 $\pm$ 0.004 &0.813 $\pm$ 0.002&0.857 $\pm$ 0.002\\

$\phantom{0}47-65\phantom{0}$ &\phantom{0}54.8&0.388 $\pm$ 0.003&0.422 $\pm$ 0.002&0.453 $\pm$ 0.002\\
$\phantom{0}65-105$&\phantom{0}80.2&0.308 $\pm$ 0.003&0.353 $\pm$ 0.002 &0.386 $\pm$ 0.001\\
$105-145$ &120.4&0.070 $\pm$ 0.001&0.096 $\pm$ 0.001&0.111 $\pm$ 0.001\\
$145-185$ &158.7& --- &0.021 $\pm$ 0.001&0.028 $\pm$ 0.001\\
  \hline
\end{tabular}
  \caption[]{The measured cross sections \seeh {} 
as a function of the \gg {} centre-of-mass energy \Wgg {} for the three data sets. Only
the statistical uncertainties, obtained after unfolding, are given.}
  \end{center}
 
\end{table}

\begin{table} 
 \label{tab:correlation}
  \begin{center} 
  \begin{tabular}{|c|c|c|c|c|c|c|c|c|}
  \hline
$\Delta$ \Wgg   (\GeV )  & 5$-$9 & 9$-$17& 17$-$31&31$-$47&47$-$65&65$-$105&105$-$145&145$-$185\\                            
  \hline
 $\phantom{0}\phantom{0}5-9\phantom{0}\phantom{0}$& 1. & & & & & & &  \\
 $\phantom{0}\phantom{0}9-17\phantom{0}$&  0.931& 1.& & & & &  &\\
 $\phantom{0}17-31\phantom{0}$& 0.815&0.939& 1.& & & &  &\\ 
 $ \phantom{0}31-47\phantom{0}$&  0.692&0.803&0.908&  1. & & &  & \\
 $ \phantom{0}47-65\phantom{0}$& 0.525 & 0.602&0.689&0.761 &1.  & &  &\\
 $\phantom{0}65-105$& 0.336&0.384 & 0.436&0.497&0.486 &1. &  & \\ 
 $ 105-145$& 0.130&0.150 & 0.166&0.186&0.190& 0.208& 1. & \\
 $ 145-185$& 0.063&0.072  & 0.077&0.080 &0.077 & 0.089 & 0.094 &1.  \\         
  \hline
\end{tabular}
  \caption[]{The correlation matrix after unfolding, for the 
 data taken at \rts = 189 \GeV .}
  \end{center}
\end{table}

\begin{table}
  \begin{center} 
\begin{tabular}{|c|c|c|c|c|c|c|c|}
  \hline

 $\Delta$ \Wgg {}(\GeV ) & Trigger & $E_{\rm{cal}}$  &$E_{\rm{lumi}}$   &$N_{\rm{part}}$ &MC stat. &Total exp.&MC model\\ 
   \hline 
  $\phantom{0}\phantom{0}5-9\phantom{0}\phantom{0}$ &     0.9    &    0.1        & $<0.1$  &  7.4 & $<0.1$  &\phantom{0}7.5 & \phantom{0}7.0 \\
  $\phantom{0}\phantom{0}9-17\phantom{0}$ &     0.9    &    0.1        & $<0.1$   &  5.0 & $<0.1$  & \phantom{0}5.1 &\phantom{0}1.2 \\  
  $\phantom{0}17-31\phantom{0}$ &     0.7    &    0.1        & $<0.1$   &  3.2 & $<0.1$  &\phantom{0}3.3 & \phantom{0}4.0 \\  
  $\phantom{0}31-47\phantom{0}$ &     0.6    &    0.2        & \phantom{0}0.2   &  2.0 & $<0.1$  &\phantom{0}2.1 & \phantom{0}6.8 \\  
  $\phantom{0}47-65\phantom{0}$ &     0.5    &    0.2        & \phantom{0}0.1   &  1.7 & $<0.1$  &\phantom{0}1.8 & \phantom{0}8.9 \\ 
  $\phantom{0}65-105$ &     0.4    &    0.2        & \phantom{0}1.1   &  1.7 & \phantom{0}\phantom{0}1.5& \phantom{0}2.6 & 10.4 \\  
  $105-145$ &     0.4    &    0.3        & \phantom{0}3.0   &  1.3 &\phantom{0}\phantom{0}8.1 &\phantom{0}8.7 &15.5 \\
  $145-185$ &     0.4    &    0.8        &\phantom{0}6.4   &  2.2 &\phantom{0}12.4 &14.1 & 27.4 \\ 
     \hline
\end{tabular}
  \caption[]{Evaluation of  systematic uncertainties due to the trigger, the analysis cuts and the Monte Carlo
  statistics. All values
  are per-cent uncertainties on the cross sections \seeh {}  and \sggh .
  The uncertainty 
  introduced by unfolding the data with PYTHIA or PHOJET is considered 
  separately in the last column.
  A further scale uncertainty of 5\%  must be added for the 
   \sggh {} cross sections, due to the two-photon luminosity function.
   }
  \end{center} 
   \label{tab:syst}
\end{table}

\begin{table}
    \label{tab:cross}
  \begin{center} 
\begin{tabular}{|c|c|c|c|c|c|c|}
  \hline
 \rts  &183 \GeV &189 \GeV &  $198$ \GeV  & all data  &  &\\ 
  \hline
$< \Wgge >(\GeV)$   &  \sgg  (nb)&  \sgg  (nb)   &  \sgg (nb)& \sgg (nb) & $\Delta  \sgge ^{\rm{exp}}$ (nb)  & $\Delta  \sgge ^{\rm{MC}}$ (nb) \\                             
 \hline
 
\phantom{0}\phantom{0}6.7 &422.6 $\pm$ 4.0&394.9 $\pm$ 0.7&398.4 $\pm$ 0.6&\,\,\,397.2 $\pm$ 0.5 & \phantom{0}30&\,\,\,$\mp$ 28\\

\phantom{0}12.3  &378.4 $\pm$ 2.8&360.2 $\pm$ 0.7&368.2 $\pm$ 0.6&\,\,\,365.2 $\pm$ 0.4 & \phantom{0}19&$\mp$ 4\\

\phantom{0}22.7 &359.8 $\pm$ 2.9&348.9 $\pm$ 0.7&358.2 $\pm$ 0.6&\,\,\,354.4 $\pm$ 0.5 & \phantom{0}12&\,\,\,$\pm$ 14\\

\phantom{0}37.8&382.1 $\pm$ 4.5 &368.4 $\pm$ 1.1&379.2 $\pm$ 0.9&\,\,\,374.8 $\pm$ 0.8 & \phantom{0}\phantom{0}8&\,\,\,$\pm$ 26\\

\phantom{0}54.8 &408.6 $\pm$ 6.4&403.0 $\pm$ 1.8&418.0 $\pm$ 1.5&\,\,\,411.5 $\pm$ 1.1 & \phantom{0}\phantom{0}7&\,\,\,$\pm$ 37\\
\phantom{0}80.2 &461.2 $\pm$ 8.6&459.5 $\pm$ 2.2&478.5 $\pm$ 1.9&\,\,\,470.3 $\pm$ 1.4 & \phantom{0}13&\,\,\,$\pm$ 49\\
120.4&496 \phantom{0}$\pm$ 19&556.7 $\pm$ 5.3&586.1 $\pm$ 4.3&\,\,\,572.0 $\pm$ 3.3 & \phantom{0}53&\,\,\,$\pm$ 89\\
158.7& --- &726\phantom{0} $\pm$ 15&738\phantom{0} $\pm$ 11 &\,\,\,734.1 $\pm$ 8.7 & 102&\,\,\,\,\,\,$\pm$ 202\\
  \hline
\end{tabular}
 \caption[]{The  \sggh~cross sections 
as a function of the average  \gg~centre-of-mass energy, $< \Wgge >$,  for the three data sets  
and for their  combination.
The statistical uncertainties, obtained after unfolding, are given for each data set.
The experimental systematic uncertainty, $\Delta  \sgge ^{\rm{exp}}$, and 
 the difference, $\Delta\sgge^{\rm MC}$,  between the average value and 
the result unfolded with PHOJET (lower sign) and with PYTHIA (upper sign) are also given.
 A further scale uncertainty of 5\%  must be added,  
 due to the two-photon luminosity function.
}
  \end{center}
 \end{table}

\begin{table}
  \begin{center} 
  \bigskip
\begin{tabular}{|l|c|c|c|c|c|c|c|}
  \hline
 Unfolding & \Wgg {} interval&  $A$ & $B$ & $\eta$ fixed&  $\epsilon$ & $\chi ^2/$d.o.f.& C.L.  \\                              
  \hline
 &\,\,\,$5-65$ \,\,\GeV&\,\,\,\,178\,\,$\pm$ \,\,5  &\,\,\,453 $\pm$ 101 & 0.358& 0.093 fixed & 5.3/3 & 0.15\\   
 &\,\,\,$5-185$ \GeV&\,\,\,\,181\,\,$\pm$ \,\,3  &\,\,\,321 $\pm$ 120 & 0.358& 0.093 fixed & \,\,55/6 & $10 ^{-9}$\\  
  &\,\,\,$5-185$ \GeV&\,\,\,\,\,\,58 $\pm$ 10  &1020 $\pm$ 146 & 0.358& 0.225 $\pm$ 0.021 & \,\,12/5 &0.04 \\
 
 PHOJET &\,\,\,$5-185$ \GeV&\,\,\,\,\,\,52 $\pm$ 11  &1201 $\pm$ 146 & 0.358& 0.221 $\pm$ 0.023 & 8.6/5 &0.12 \\ 
 PYTHIA &\,\,\,$5-185$ \GeV&\,\,\,\,\,\,63 $\pm$ 10  &\,\,\,842 $\pm$ 146 & 0.358& 0.228 $\pm$ 0.018 & \,\,19/5 &0.002 \\

 \hline
 &$17-105$ \GeV&\,\,\,165 $\pm$ 21&---&---&0.116 $\pm$ 0.016&4.3/2 & 0.12 \\ 
 &$31-185$ \GeV&\,\,\,113 $\pm$ 19&---&---&0.163 $\pm$ 0.021&3.4/3 & 0.33 \\
 &$47-185$ \GeV&\,\,\,\,\,\,81 $\pm$ 23&---&---&0.202 $\pm$ 0.035&1.5/2 & 0.48 \\

  \hline
\end{tabular}
  \caption[]{Fits to the total cross sections listed in Table 4       
of the form 
 \sgg 
$= A \, s^{\epsilon} \, + \, B \, s^{-\eta}$\cite{DL},
where $s=W_{\gamma \gamma }^2$ . PHOJET and  PYTHIA   indicates that  only this Monte
 Carlo is used to unfold the data. In all other cases the average unfolding result
 of the two generators is used.
 The statistical and experimental
uncertainties and the correlation matrix between the data points  are used. 
The fitted parameters are strongly correlated. The second set 
of fits  evaluates only the increase of \sgg {} with $s$, {\it i.e.} the 
 Pomeron part of the fit. The values of the $\chi ^2$ and the corresponding confidence level
 are  given.}
  \end{center} 
  \label{tab:fit}
\end{table}

\newpage


\end{document}